\documentclass[english,aps,preprint,nofootinbib]{revtex4-1}
\usepackage[T1]{fontenc}
\usepackage[latin9]{inputenc}
\usepackage{booktabs}
\usepackage{amsmath}
\usepackage{amssymb}
\usepackage{graphicx}
\usepackage{esint}

\makeatletter

\providecommand{\tabularnewline}{\\}

 \@ifundefined{textcolor}{}
 {%
   \definecolor{BLACK}{gray}{0}
   \definecolor{WHITE}{gray}{1}
   \definecolor{RED}{rgb}{1,0,0}
   \definecolor{GREEN}{rgb}{0,1,0}
   \definecolor{BLUE}{rgb}{0,0,1}
   \definecolor{CYAN}{cmyk}{1,0,0,0}
   \definecolor{MAGENTA}{cmyk}{0,1,0,0}
   \definecolor{YELLOW}{cmyk}{0,0,1,0}
 }

\usepackage{babel}

\usepackage{colortbl}

\definecolor{gray}{rgb}{0.8, 0.8,0.8}

\makeatother

\usepackage{babel}
\begin{document}

\global\long\def\mpl{m_{\mathrm{Pl}}}
 \global\long\def\e{\mathrm{end}}
 \global\long\def\k{\kappa}
 \global\long\def\f{\varphi}
 \global\long\def\gy{h}

\title{Does Planck really rule out monomial inflation?}

\author{Kari Enqvist}

\author{Mindaugas Kar\v{c}iauskas}

\affiliation{Physics Department and Helsinki Institute of Physics\\
 PO Box 64, FIN-00014 University of Helsinki}
\begin{abstract}
We consider the modifications of monomial chaotic inflation models
due to radiative corrections induced by inflaton couplings to bosons
and/or fermions necessary for reheating. To the lowest order, ignoring
gravitational corrections and treating the inflaton as a classical
background field, they are of the Coleman-Weinberg type and parametrized
by the renormalization scale $\mu$. In cosmology, there are not enough
measurements to fix $\mu$ so that we end up with a family of models,
each having a slightly different slope of the potential. We demonstrate
by explicit calculation that within the family of chaotic $\phi^{2}$
models, some may be ruled out by Planck whereas some remain perfectly
viable. In contrast, radiative corrections do not seem to help chaotic
$\phi^{4}$ models to meet the Planck constraints. 
\end{abstract}
\maketitle

\section{Introduction}

Planck data has famously been used to constrain single-field inflaton
models, such as large-field models with a monomial potential $V\sim\phi^{n}$.
Such models can be considered as effective particle physics theories
with heavy degrees of freedom and/or interactions with other fields
integrated out. From such a point of view one may justify e.g. neglecting
the running of $\lambda$ in $\lambda\phi^{4}$ model and treat it
as an effective constant. Adopting such an approach, Planck rules
out chaotic $\lambda\phi^{4}$ models while $m^{2}\phi^{2}$ models
may still be allowed, albeit marginally.

However, we should like to point out that inflaton decay is an essential
part of any inflationary scenario that cannot be integrated out. Thus,
any model must be augmented by a mechanism that brings inflation to
an end and reheats the universe. As is well known, this means adding
interaction terms to a model so that to lowest order the potential
in monomial inflation would read like 
\begin{equation}
V\left(\phi,\chi,\psi\right)=\frac{1}{2}\lambda_{\mathrm{b}}\mpl^{4}\left(\frac{\phi}{\mpl}\right)^{n}-\frac{1}{2}g_{\mathrm{b}}^{2}\phi^{2}\chi^{2}-\gy_{\mathrm{b}}\phi\bar{\psi}\psi+\ldots\label{V-def}
\end{equation}
where $\chi$ is some bosonic and $\psi$ fermionic field, and the
subscript `$\mathrm{b}$' denotes bare coupling constant values. Dots
represent other terms, such as mass terms for the fields $\chi$ and
$\psi$. These interactions will then generate through loop corrections
operators that modify the effective inflaton potential. Here we focus
on the minimal modifications only. We take $\chi$ and $\psi$ to
be quantum fields while the inflaton $\phi$ is treated as a classical
background field. In particular, this means neglecting the inflaton
loops. Moreover, we do not consider quantum corrections in curved
space background, which can induce additional curvature-dependent
terms \cite{Elizalde(1995)Veff} into the potential \eqref{V-def}.
The issue with the curved space corrections is not just the vacuum
structure but there may also arise corrections to the slow-roll equations
of the mean field \cite{Herranen(2013)Qcorr}. To circumvent these
complications, and for the sake of clarity, we view the first term
in \eqref{V-def} as the effective energy of an order parameter that
is the source of the Friedmann equation after integrating out the
additional gravitational effects.

To lowest order the modifications are then of the conventional Coleman-Weinberg
type. The properly regulated effective inflaton model would thus read
\begin{eqnarray}
V_{\mathrm{eff}}\left(\phi\right) & = & \frac{1}{2}\lambda\mpl^{4}\left(\frac{\phi}{\mpl}\right)^{n}+\frac{g^{4}-4\gy^{4}}{64\pi^{2}}\phi^{4}\ln\frac{\phi^{2}}{\mu^{2}},\label{V-eff}
\end{eqnarray}
where $g$ and $h$ are renormalized coupling constants, and by virtue
of the classical nature of $\phi$, $\lambda=\lambda_{\mathrm{b}}$.
Renormalizability requires $n\le4$ and we have assumed $g\phi\gg m_{\chi}$
and $\gy\phi\gg m_{\psi}$, where $m_{\chi}$ and $m_{\psi}$ are
masses of $\chi$ and $\psi$ fields respectively and these terms
are included in the part of the potential denoted by ellipsis in Eq.~\eqref{V-def}.
For self-consistency of the model, radiative corrections to the potential
must be taken into account when estimating the number of e-folds and
values of the slow-roll parameters. This is the purpose of the present
paper.

The importance of radiative corrections for chaotic inflation was
already pointed out by Senoguz and Shafi in \cite{NeferSenoguz(2008)Qcorr}.
They considered only fermionic contributions, but more importantly,
they chose the renormalization scale as $\mu=hM_{P}$. The renormalization
scale is of course arbitrary; there is no \textquotedbl{}natural\textquotedbl{}
scale $\mu$ except in the technical sense of minimizing higher order
corrections. In particle physics, one trades $\mu$ with a measured
value of some physical amplitude; this is the act of \textquotedbl{}normalization\textquotedbl{}.
For instance, one could measure a $2\to2$ scattering amplitude at
some fixed external momenta ${\bf p}$ (and most conveniently at the
symmetric point with all the momenta equal) to extract the coupling
constant at the renormalization point $p^{2}=\mu^{2}$. After that,
the truncated perturbative expression yields the running of the amplitude
(or coupling) as the response to the scaling of the external momenta.
For inflationary models, the situation is trickier. For instance,
the physical inflaton mass could be defined as the pole mass $m_{{\rm phys}}=m(p^{2}=m_{{\rm phys}}^{2})$,
where $m$ is the bare mass, which is a parameter in the potential.
Unfortunately, there are no prospects for measuring the physical inflaton
mass independently of cosmological observations. Thus the question
is, what exactly is the meaning of the potential parameters that can
be constrained by CMB observations - and which models are truly ruled
out?

As far as the Planck constraints are concerned, we shall point out
that different renormalization points correspond to physically different
models of inflation in that they lead to different predictions for
the observables $n_{s}$ and $r$. The models are different in the
sense that the shape of the potential at some fixed $\phi$ is different
for different renormalization points; likewise, given identical initial
conditions, they would yield a different number of e-folds. Alternatively,
by fixing, say, the value of the spectral index, one would obtain
a family of models with the same $n_{s}$ but with different physical
model parameters $m$ (or $\lambda$) and $g$ or $h$. As we shall
show, for the $\phi^{2}$ potential some of the models in this family
are ruled out while some remain perfectly viable. We should emphasize
that we consider only cases where the radiative corrections are always
small. For the $\phi^{4}$ potential this restricts the possible modifications
so that this class of models is always ruled out by the Planck data.

Here we differ from the approach of Senoguz and Shafi \cite{NeferSenoguz(2008)Qcorr}
who fix the renormalization point and claim that the fixing does not
affect physics. This is of course true in the sense that observables,
such as scattering cross sections, are not affected. However, cosmological
constraints on model parameters -- which are not observables -- very
much depend on at which scale those parameters are being defined.

\section{Planck constraints on large-field monomial inflation}

Let us provide a brief summary of the relevant Planck results and
constraints. In single field inflation models the value of the energy
density at every space-time position $\left(\boldsymbol{x},t\right)$
is determined completely by the value of the inflaton field $\phi\left(\boldsymbol{x},t\right)$.
Hence, $\phi\left(\boldsymbol{x},t\right)$ also determines the time
shift between the flat and uniform energy density hypersurfaces, where
the perturbation spectrum is being computed. It is given by \cite{Lyth_Liddle(2009)book}
\begin{equation}
\mathcal{P}_{\zeta}\left(k\right)=\frac{1}{24\pi^{2}\mpl^{4}}\left.\frac{V\left(\phi\right)}{\epsilon}\right|_{k}.\label{Pz}
\end{equation}
The perturbation amplitude $\mathcal{P}_{\zeta}\left(k_{*}\right)$
at the the pivot scale $k_{*}=0.05\mathrm{\, Mpc^{-1}}$ is measured
by Planck as \cite{Planck_XVI} 
\begin{equation}
\mathcal{P}_{\zeta}\left(k_{*}\right)=2.20\times10^{-9}.\label{normalisation}
\end{equation}

Large-field monomial inflaton models are slow-roll models. The spectral
index is given as usual by 
\begin{equation}
n_{\mathrm{s}}-1=-6\epsilon+2\eta,\label{ns-expr}
\end{equation}
and the running of the spectral index 
\begin{equation}
n'\equiv\mathrm{d}n_{\mathrm{s}}/\mathrm{d}\ln k=-24\epsilon^{2}+16\epsilon\eta-2\xi,\label{sp-running}
\end{equation}
where slow-roll parameters are defined as 
\begin{equation}
\epsilon\equiv\frac{\mpl^{2}}{2}\left(\frac{V'}{V}\right)^{2},\,\eta\equiv\mpl^{2}\frac{V''}{V},\,\xi\equiv\mpl^{4}\frac{V'V'''}{V^{2}},\label{slow-roll-prms}
\end{equation}
and $V\equiv V\left(\phi\right)$ is the potential of the inflaton.
Primes denote derivatives with respect to $\phi$. During inflation
all slow-roll parameters are small, $\epsilon,\,\eta,\,\xi\ll1$. 

Tensor-to-scalar ratio obeys the relation 
\begin{equation}
r=16\epsilon.\label{r-expr}
\end{equation}
Hence, measuring primordial gravitational waves would allow a direct
determination of the energy scale of inflation from Eqs.~\eqref{Pz}
and \eqref{normalisation}.

Currently measured value of $n_{\mathrm{s}}$ is \cite{Planck_XXII}
(assuming no running and tenser modes) 
\begin{equation}
n_{\mathrm{s}}=0.9603\pm0.0073.
\end{equation}
Allowing for spectral running Planck constraints give 
\begin{eqnarray}
n_{\mathrm{s}} & = & 0.9630\pm0.0065,\\
n' & = & -0.013\pm0009
\end{eqnarray}
at $68\%\:\mathrm{CL}$ at the decorrelation pivot scale $k_{*}^{\mathrm{dec}}=0.038\,\mathrm{Mpc}^{-1}$.
As noted in Ref.~\cite{Planck_XXII} the value of $\xi$ derived
from this measurement is still compatible with zero at $95\%$ CL.

The contamination of the primordial B-mode spectrum mainly by the
gravitational lensing signal sets the lowest bound on the value of
$r$ which one might hope to ever achieve if not detected. This limit
is $10^{-4}$ or so \cite{Knox(2002)rLimit}. There is, however, a
class of inflationary models which produce larger $r$ than this limit.
Indeed, as emphasized in the Introduction, current observational bounds
on $r$ decreased to the level where it becomes possible to falsify
some of these models. The most stringent constraints on $r$ are derived
from the recent Planck satellite results \cite{Planck_XXII} 
\begin{eqnarray}
n_{\mathrm{s}} & = & 0.9624\pm0.0075\\
r & < & 0.12
\end{eqnarray}
at $95\%\:\mathrm{CL}$ and at a pivot scale $k_{*}=0.002\,\mathrm{Mpc}^{-1}$.

The Planck team used these results to constrain some of the inflationary
models in Ref.~\cite{Planck_XXII}. Assuming a simple tree-level
potential corresponding to the first term of Eq.~\eqref{V-def},
they showed that such a low value of $r$ is incompatible with large-field
models with a monomial potential $\sim\phi^{n}$, where $n=3\;\mathrm{and}\;4$
and only marginally compatible with the $n=2$ model. They also constrained
the linear model as well as a model with $n=2/3$, both of which are
not amenable to conventional perturbation theory. We do not consider
such models here.

\section{The Effect of Radiative Corrections}

\subsection{Chaotic Inflation with Radiative Corrections}

To simplify the expressions let us rewrite the inflaton potential
in Eq.~\eqref{V-eff} as 
\begin{equation}
V_{\mathrm{eff}}\left(\phi\right)=\frac{1}{2}\lambda\mpl^{4}\left(\f^{n}+\k\f^{4}\ln\frac{\varphi}{\mu}\right),\label{Vtilde}
\end{equation}
where $\varphi$ and $\kappa$ are defined as 
\begin{eqnarray}
\f & \equiv & \frac{\phi}{\mpl}
\end{eqnarray}
and

\begin{equation}
\k\equiv\frac{g^{4}-4\gy^{4}}{16\pi^{2}\lambda}.
\end{equation}
In this expression the renormalization scale $\mu$ is given in units
of the Planck mass. One can immediately notice that the potential
can become unbounded from bellow when the second term in Eq.~\eqref{Vtilde}
is negative and dominates. In this regime, however, higher order loop
corrections become important and should be included in Eq.~\eqref{Vtilde}.
As we mainly constrain ourselves within the regime of small radiative
corrections, this apparent instability does not have to concern us. 

Looking at Eq.~\eqref{Vtilde} it should be clear that radiative
corrections change both the slope and the curvature of the potential.
Due to Eqs.~\eqref{ns-expr}, \eqref{sp-running} and \eqref{r-expr}
observables of CMB, such as the spectral index $n_{\mathrm{s}}$,
its running $n'$, and tensor-to-scalar ratio $r$ are also modified
from their tree level values. Using Eq.~\eqref{slow-roll-prms} one
can easily compute the radiatively corrected slow-roll parameters
as 
\begin{eqnarray}
\epsilon & = & \frac{K(\varphi)^{2}}{2\f^{2}}\left[n+\k\f^{4-n}\left(1+4\ln\frac{\varphi}{\mu}\right)\right]^{2},\label{eps}\\
\eta & = & \frac{K(\varphi)}{\f^{2}}\left[n\left(n-1\right)+\k\f^{4-n}\left(7+12\ln\frac{\varphi}{\mu}\right)\right]\,,\label{eta}\\
\xi & = & \frac{K(\varphi)^{2}}{\varphi^{4}}\left[n+\kappa\varphi^{4-n}\left(1+4\ln\frac{\varphi}{\mu}\right)\right]\left[n\left(n-1\right)\left(n-2\right)+\kappa\varphi^{4-n}\left(14+46\ln\frac{\varphi}{\mu}\right)\right]\label{xi}
\end{eqnarray}
where for brevity we defined 
\begin{equation}
K^{-1}(\varphi)\equiv1+\k\f^{4-n}\ln\frac{\varphi}{\mu}.
\end{equation}
Knowing the values of these parameters a couple of e-folds after the
pivot scale $k_{*}$ exits the horizon, it is easy to compute $n_{\mathrm{s}}\left(n,\kappa,\mu\right)$,
$n'\left(n,\kappa,\mu\right)$ and $r\left(n,\kappa,\mu\right)$ using
Eqs.~\eqref{ns-expr}, \eqref{sp-running} and \eqref{r-expr} respectively.
However, in contrast to the tree level potential, the values of these
parameters are determined not only by the power $n$ but also by the
strength of the couplings of the inflaton to other fields $\kappa$,
as well as by the renormalization scale $\mu$.

To find numerical values of slow-roll parameters in Eqs.~\eqref{eps}
- \eqref{xi} we need to solve the system of three coupled equations.
The first equation is given by the end of inflation condition. Inflation
terminates when the slow-roll parameter $\epsilon$ becomes of order
one. Hence, we define the inflaton value at the end of inflation $\varphi_{\e}$
as 
\begin{eqnarray}
\epsilon\left(\varphi_{\e},n,\kappa,\mu\right) & = & 1.\label{eq1}
\end{eqnarray}
The number of e-folds of inflation from the time when a mode $k$
leaves the horizon to the end of inflation is given by $N_{k}\equiv\int_{t_{k}}^{t_{\e}}H\mathrm{d}t=\mpl^{-2}\int_{\varphi_{\e}}^{\varphi_{k}}V/V'\mathrm{d}\varphi$,
where we used the slow-roll result $3H\dot{\varphi}\simeq-V'$. Plugging
Eq.~\eqref{slow-roll-prms} into this result we can write for the
pivot scale $k_{*}$ 
\begin{equation}
\int_{\varphi_{*}}^{\varphi_{\e}}\frac{\mathrm{d}\varphi}{\sqrt{2\epsilon\left(\varphi,n,\kappa,\mu\right)}}=-N_{*},\label{eq2}
\end{equation}
where $N_{*}\equiv N_{k_{*}}$. One also has to make sure that the
solution gives the curvature perturbation amplitude in Eq.~\eqref{Pz}
consistent with the observed value in Eq.~\eqref{normalisation}.
This provides a third equation to compute the (renormalised) inflaton
self-coupling constant $\lambda$ in Eq.~\eqref{Vtilde} 
\begin{equation}
\lambda=24\pi^{2}\mathcal{P}_{*}\frac{\left[n+\kappa\varphi_{*}^{4-n}\left(1+4\ln\frac{\varphi_{*}}{\mu}\right)\right]^{2}}{\varphi_{*}^{2+n}\left[1+\kappa\varphi_{*}^{4-n}\ln\frac{\varphi_{*}}{\mu}\right]^{3}}.\label{eq3}
\end{equation}

The value of $N_{*}$ in Eq.~\eqref{eq2} depends on the temperature
of reheating $T_{\mathrm{reh}}\propto\rho_{\mathrm{reh}}^{1/4}$ and
is given by \cite{Liddle(2003)efolds} 
\begin{equation}
N_{*}=68.5+\frac{1}{2}\ln\frac{V_{*}}{\mpl^{4}}-\frac{1}{3}\ln\frac{V_{\e}}{\mpl^{4}}+\frac{1}{12}\ln\frac{\rho_{\mathrm{reh}}}{\mpl^{4}},\label{Ns}
\end{equation}
where $V_{*}\equiv V_{\mathrm{eff}}\left(\phi_{*}\right)$, $V_{\e}\equiv V_{\mathrm{eff}}\left(\phi_{\e}\right)$
and we also used the value of the present day Hubble constant $H_{0}=67.04$
km/s/Mpc \cite{Planck_XVI}. We assume in this work that the inflaton
decays through a process of perturbative decay. As $N_{*}$ depends
only on the temperature of reheating, non-perturbative effects will
not in general change the results much. In some cases, however, these
effects could change the thermal history of the universe \cite{Kofman(1997)preh}.
Such cases should be treated separately. For the perturbative decay,
on the other hand, we can write 
\begin{equation}
\rho_{\mathrm{reh}}\simeq3\mpl^{2}\Gamma^{2},
\end{equation}
where $\Gamma$ is the decay rate of the inflaton.

Scalar spectral index $n_{s}$, spectral running $n'$ and tensor-to-scalar
ratio $r$ can be calculated by solving Eqs.~\eqref{eq1} - \eqref{eq3}
for $\varphi_{*}$ and plugging the result into Eqs.~\eqref{ns-expr},
\eqref{sp-running} and \eqref{r-expr}. Unfortunately it is impossible
to solve them analytically. Therefore, one has to resort to numerical
methods. As we scan over different values of $\kappa$ to find solutions
of Eqs.~\eqref{eq1} - \eqref{eq3}, we must also make sure that
the universe reheats well before the start of Big Bang Nucleosynthesis,
which happens at $T_{\mathrm{BBN}}\sim1\:\mathrm{MeV}$. Thus the
minimum value of $\left|\kappa\right|$ must be constrained to give
$T_{\mathrm{reh}}\sim\sqrt{\mpl\Gamma}>T_{\mathrm{BBN}}$. This bound
however, is many orders of magnitude below the values of interest
for this work.
\begin{figure}
\centering{}\includegraphics[scale=0.5]{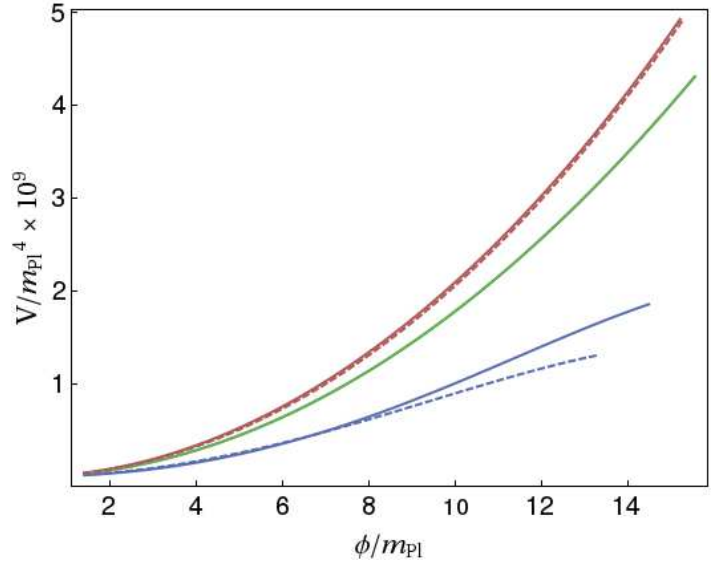}\caption{\label{fig:V} Inflationary parts of $m^{2}\phi^{2}$ potentials.
The middle, solid green curve correspond to the potential with $N_{*}=60$
and $\kappa=0$ (denoted by the large black dot in Figure~\ref{fig:Rezs}).
Lower blue curves correspond to inflaton potentials denoted by number
1 in Figure~\ref{fig:Rezs}, while top red curves correspond to number
6 in that Figure.}
\end{figure}

\subsection{The Case of $n=2$}

\begin{figure}
\centering{}\includegraphics[scale=0.5]{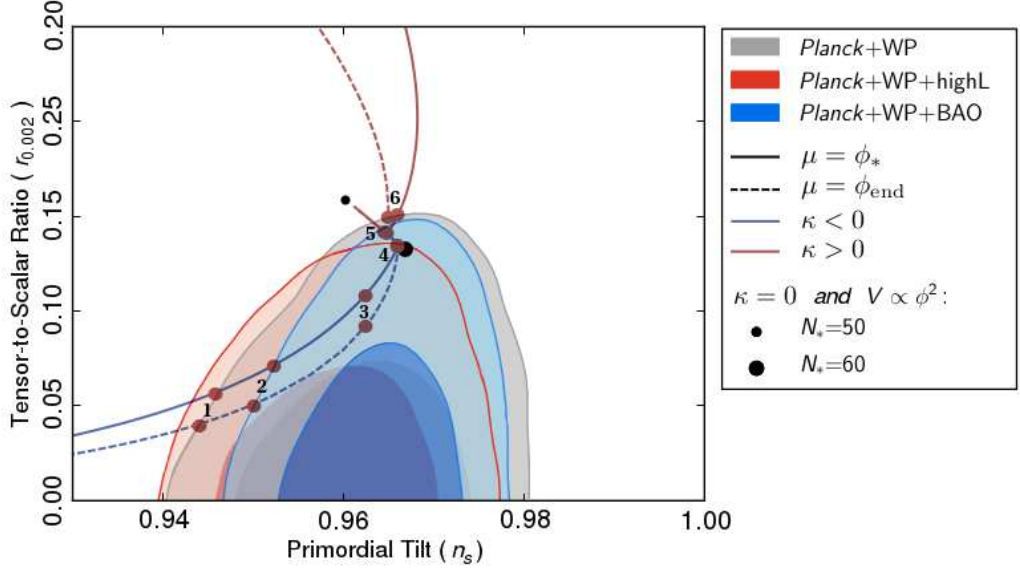}\caption{\label{fig:Rezs}Planck constraints on the spectral index $n_{\mathrm{s}}$
and tensor-to-scalar ratio $r$ \cite{Planck_XXII}. Black dots correspond
to the $\phi^{2}$ model with $\kappa=0$. Solid and dashed curves
correspond to normalization scales $\mu=\phi_{*}$ and $\mu=\phi_{\e}$
respectively. Blue curves show the effect of radiative corrections
when the $\phi\rightarrow\bar{\psi}\psi$ decay channel dominates
($\kappa<0$) and red curves when the $\phi\rightarrow\chi\chi$ channel
dominates ($\kappa>0$). The values of parameters in Table~\ref{tab:rez}
are denoted by numbers. }
\end{figure}
\begin{table}
\begin{centering}
\begin{tabular}{cccccccc}
\multicolumn{8}{c}{$\mu=\phi_{*}$}\tabularnewline
\midrule 
 & {\footnotesize{{1}}}  & {\footnotesize{{2}}}  & {\footnotesize{{3}}}  & {\footnotesize{{4}}}  &  & {\footnotesize{{5}}}  & {\footnotesize{{6}}}\tabularnewline
\midrule
\midrule 
\addlinespace[-1mm]
{\footnotesize{{$\kappa$ }}}  & {\footnotesize{{$-3.7\times10^{-3}$}}}  & {\footnotesize{{$-2.8\times10^{-3}$}}}  & {\footnotesize{{$-1.0\times10^{-3}$}}}  & {\footnotesize{{$-2.0\times10^{-5}$ }}}  & {\footnotesize{{$0$}}}  & {\footnotesize{{$3.0\times10^{-5}$ }}}  & {\footnotesize{{$4.0\times10^{-4}$}}}\tabularnewline\addlinespace[1mm]
\addlinespace[-2mm]
\rowcolor{gray}{\footnotesize{{$n_{\mathrm{s}}$}}}  & {\footnotesize{{$0.9459$}}}  & {\footnotesize{{$0.9524$}}}  & {\footnotesize{{$0.9625$}}}  & {\footnotesize{{$0.9661$ }}}  & {\footnotesize{{$0.9670$ }}}  & {\footnotesize{{$0.9645$ }}}  & {\footnotesize{{$0.9661$}}}\tabularnewline
\addlinespace[-2mm]
\rowcolor{gray}{\footnotesize{{$r$}}}  & {\footnotesize{{$0.0570$}}}  & {\footnotesize{{$0.0716$}}}  & {\footnotesize{{$0.1089$}}}  & {\footnotesize{{$0.1349$ }}}  & {\footnotesize{{$0.132$ }}}  & {\footnotesize{{$0.1417$ }}}  & {\footnotesize{{$0.1513$}}}\tabularnewline
\addlinespace[-2mm]
\rowcolor{gray}{\footnotesize{{$n'\times10^{4}$}}}  & {\footnotesize{{$-6.345$}}}  & {\footnotesize{{$-7.197$}}}  & {\footnotesize{{$-7.122$}}}  & {\footnotesize{{$-5.774$}}}  & {\footnotesize{{$-5.464$ }}}  & {\footnotesize{{$-6.140$}}}  & {\footnotesize{{$-5.166$}}}\tabularnewline
{\footnotesize{{$h,\, g\times10^{3}$ }}}  & {\footnotesize{{$1.268$}}}  & {\footnotesize{{$1.242$}}}  & {\footnotesize{{$1.050$}}}  & {\footnotesize{{$0.414$}}}  & {\footnotesize{{$0$}}}  & {\footnotesize{{$0.662$ }}}  & {\footnotesize{{$1.280$}}}\tabularnewline
\addlinespace[-2mm]
{\footnotesize{{$m$$\left(10^{13}\:\mathrm{GeV}\right)$}}}  & {\footnotesize{{$1.009$}}}  & {\footnotesize{{$1.113$}}}  & {\footnotesize{{$1.332$}}}  & {\footnotesize{{$1.464$ }}}  & {\footnotesize{{$1.01$ }}}  & {\footnotesize{{$1.530$ }}}  & {\footnotesize{{$1.566$}}}\tabularnewline
\addlinespace[-2mm]
{\footnotesize{{$\phi_{*}/\mpl$}}}  & {\footnotesize{{$14.49$}}}  & {\footnotesize{{$14.72$}}}  & {\footnotesize{{$15.17$}}}  & {\footnotesize{{$15.37$ }}}  & {\footnotesize{{$15.56$ }}}  & {\footnotesize{{$15.08$ }}}  & {\footnotesize{{$15.22$}}}\tabularnewline
\addlinespace[-2mm]
{\footnotesize{{$N_{*}$}}}  & {\footnotesize{{$58.67$}}}  & {\footnotesize{{$58.73$}}}  & {\footnotesize{{$58.81$}}}  & {\footnotesize{{$58.56$}}}  & {\footnotesize{{$60$ }}}  & {\footnotesize{{$56.23$ }}}  & {\footnotesize{{$56.74$}}}\tabularnewline
\bottomrule
\end{tabular}
\par\end{centering}

\begin{centering}
{\scriptsize{}}%
\begin{tabular}{cccccccc}
\addlinespace[5mm]
\multicolumn{8}{c}{$\mu=\phi_{\e}$}\tabularnewline
\midrule 
 & {\footnotesize{{1}}}  & {\footnotesize{{2}}}  & {\footnotesize{{3}}}  & {\footnotesize{{4}}}  &  & {\footnotesize{{5}}}  & {\footnotesize{{6}}}\tabularnewline
\midrule
\midrule 
\addlinespace[-1mm]
{\footnotesize{{$\kappa$ }}}  & {\footnotesize{{$-7.5\times10^{-4}$ }}}  & {\footnotesize{{$-6.4\times10^{-4}$ }}}  & {\footnotesize{{$-3.0\times10^{-4}$ }}}  & {\footnotesize{{$-2.0\times10^{-6}$}}}  & {\footnotesize{{$0$ }}}  & {\footnotesize{{$8.0\times10^{-6}$ }}}  & {\footnotesize{{$7.0\times10^{-5}$}}}\tabularnewline
\addlinespace[1mm]
\addlinespace[-2mm]
\rowcolor{gray}{\footnotesize{$n_{s}$}} & {\footnotesize{{$0.9441$}}}  & {\footnotesize{{$0.9501$}}}  & {\footnotesize{{$0.9626$}}}  & {\footnotesize{{$0.9660$ }}}  & {\footnotesize{{$0.9670$ }}}  & {\footnotesize{{$0.9647$ }}}  & {\footnotesize{{$0.9650$ }}}\tabularnewline
\addlinespace[-2mm]
\rowcolor{gray}{\footnotesize{{$r$ }}}  & {\footnotesize{{$0.0403$}}}  & {\footnotesize{{$0.0509$}}}  & {\footnotesize{{$0.0926$}}}  & {\footnotesize{{$0.1356$ }}}  & {\footnotesize{{$0.132$ }}}  & {\footnotesize{{$0.1425$ }}}  & {\footnotesize{{$0.1503$ }}}\tabularnewline
\addlinespace[-2mm]
\rowcolor{gray}{\footnotesize{{$n'\times10^{4}$}}}  & {\footnotesize{{$3.627$}}}  & {\footnotesize{{$1.876$}}}  & {\footnotesize{{$-2.977$}}}  & {\footnotesize{{$-5.757$}}}  & {\footnotesize{{$-5.464$ }}}  & {\footnotesize{{$-6.308$ }}}  & {\footnotesize{{$-6.630$ }}}\tabularnewline
\addlinespace
{\footnotesize{{$h,\, g\times10^{4}$ }}}  & {\footnotesize{{$8.885$}}}  & {\footnotesize{{$8.856$}}}  & {\footnotesize{{$7.918$}}}  & {\footnotesize{{$2.334$ }}}  & {\footnotesize{{$0$ }}}  & {\footnotesize{{$4.764$ }}}  & {\footnotesize{{$8.182$ }}}\tabularnewline
\addlinespace[-2mm]
{\footnotesize{{$m\left(10^{13}\,\mathrm{GeV}\right)$ }}}  & {\footnotesize{{$1.101$}}}  & {\footnotesize{{$1.184$}}}  & {\footnotesize{{$1.383$}}}  & {\footnotesize{{$1.471$}}}  & {\footnotesize{{$1.01$ }}}  & {\footnotesize{{$1.533$ }}}  & {\footnotesize{{$1.528$ }}}\tabularnewline
\addlinespace[-2mm]
{\footnotesize{{$\phi_{*}/\mpl$ }}}  & {\footnotesize{{$13.36$}}}  & {\footnotesize{{$13.67$}}}  & {\footnotesize{{$14.62$}}}  & {\footnotesize{{$15.34$ }}}  & {\footnotesize{{$15.56$ }}}  & {\footnotesize{{$15.06$ }}}  & {\footnotesize{{$15.25$ }}}\tabularnewline
\addlinespace[-2mm]
{\footnotesize{{$N_{*}$}}}  & {\footnotesize{{$58.35$ }}}  & {\footnotesize{{$58.43$}}}  & {\footnotesize{{$58.62$}}}  & {\footnotesize{{$58.37$}}}  & {\footnotesize{{$60$ }}}  & {\footnotesize{{$56.06$ }}}  & {\footnotesize{{$56.45$ }}}\tabularnewline
\bottomrule
\end{tabular} 
\par\end{centering}

\caption{\label{tab:rez}Numerical values of some parameters for two choices
of renormalization scale $\mu$ and different values of inflaton interaction
strength $\kappa$. Each value of $\kappa$ correspond to points in
Figures~\ref{fig:Rezs} and \ref{fig:dn} marked by numbers, which
are also shown in the top rows of these tables. Smallest $\left|\kappa\right|$'s
are chosen to lie at the turnover where $\mu=\phi_{*}$ and $\mu=\phi_{\e}$
curves converge on the line joining $N_{*}=50$ and $N_{*}=60$ points.
Largest values of $\left|\kappa\right|$ are chosen to lie on the
border of $95\%$ CL of ``Planck+WP'' contour in Figure~\ref{fig:Rezs}.
The `$h$, $g$' row displays the values of the coupling constant
$h$ for $\kappa<0$ and $g$ for $\kappa>0$.}
\end{table}
For the quadratic monomial potential with $n=2$ we can write 
\begin{equation}
\lambda=\frac{m^{2}}{\mpl^{2}},
\end{equation}
where $m^{2}$ is the \emph{renormalised} mass of the inflaton. The
inflaton decay rate to bosons $\phi\rightarrow\chi\chi$ is given
by 
\begin{equation}
\Gamma_{\chi}=\frac{g^{4}\sigma^{2}}{8\pi m},
\end{equation}
where $\sigma$ is the inflaton vacuum expectation value. We assume
$\sigma\sim m$ for simplicity. Since $N_{*}$ depends only logarithmically
on $\sigma$, the change of the latter will not have a major effect.
While the decay rate to fermions $\phi\rightarrow\psi\bar{\psi}$
is given by 
\begin{equation}
\Gamma_{\psi}=\frac{\gy^{2}m}{8\pi}.
\end{equation}
The total decay rate of the inflaton is a sum of the two $\Gamma=\Gamma_{\chi}+\Gamma_{\psi}$.
However, we consider only the cases when either $\Gamma_{\chi}$ or
$\Gamma_{\psi}$ dominates.

\begin{figure}
\centering{}\includegraphics[scale=0.5]{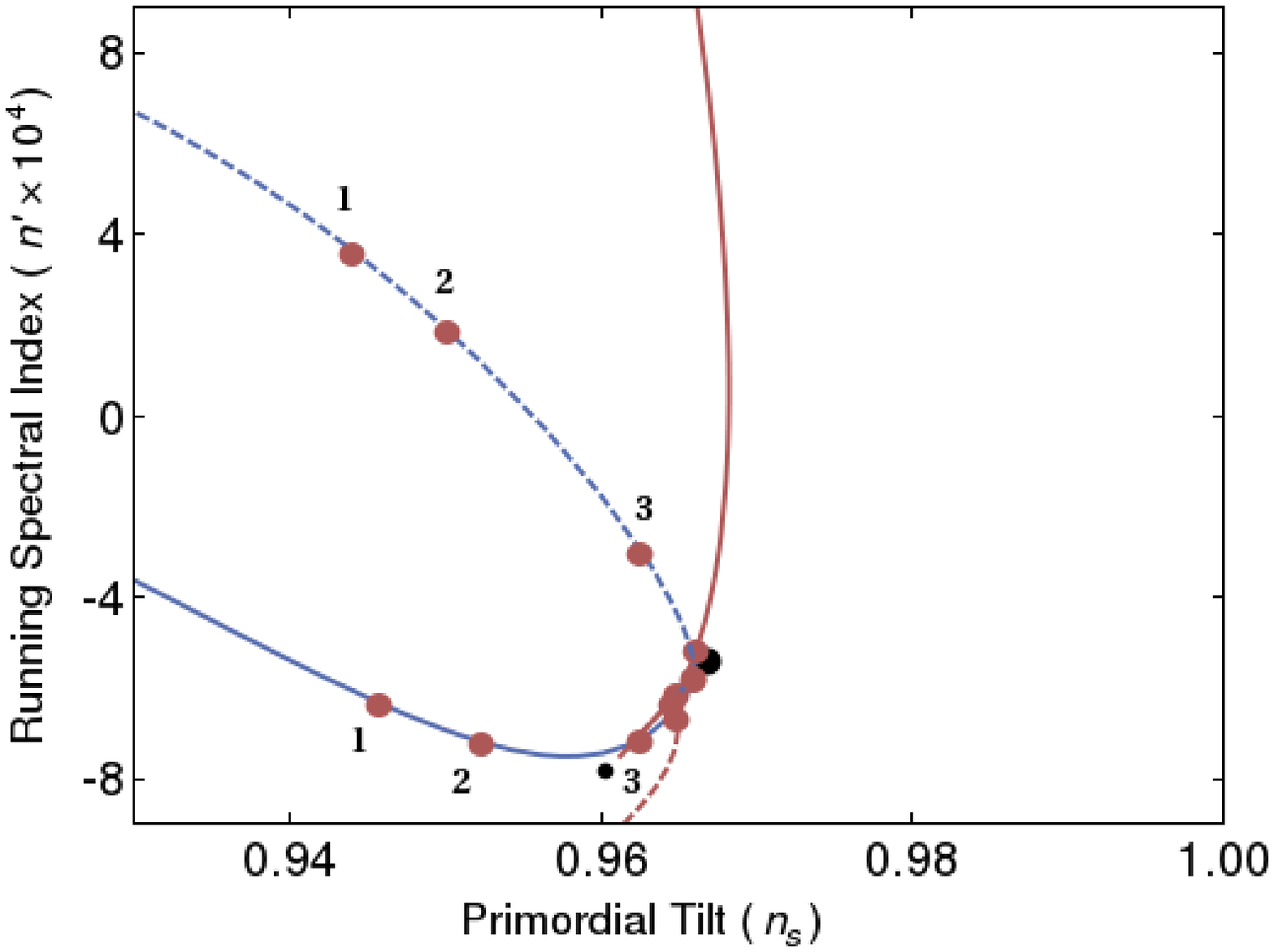} \caption{\label{fig:dn}Plot of spectral index $n_{s}$ versus spectral running
$n'$. Numbers correspond to the same points in Figure~\ref{fig:Rezs}
and their numerical values are given in Table~\ref{tab:rez}. The
notation is the same as in Figure~\ref{fig:Rezs}: black dots correspond
to $\kappa=0$ as in Ref.~\cite{Planck_XXII}, blue curves to $\kappa<0$
and red ones to $\kappa>0$, solid curves correspond to $\mu=\phi_{*}$
and dashed ones to $\mu=\phi_{\e}$.}
\end{figure}

We solve Eqs.~\eqref{eq1} - \eqref{eq3} for two choices of renormalization
scale $\mu$. In the first case $\mu$ is such that radiative corrections
vanish when the pivot scale exits the horizon, i.e. $\mu=\phi_{*}$.
In effect, this then corresponds to a \textquotedbl{}pure\textquotedbl{}
$m^{2}\phi^{2}$ model valid at the pivot scale only. The other extreme
is the case with $\mu=\phi_{\e}$, where the \textquotedbl{}pure\textquotedbl{}
$m^{2}\phi^{2}$ model is valid at the end of inflation only. The
effect of radiative corrections on the inflationary part of the tree
level potential is illustrated in Figure~\ref{fig:V}.

As Eqs.~\eqref{eq1} - \eqref{eq3} cannot be solved analytically,
we list out for illustrative purposes some numerical results in Table~\ref{tab:rez}.
We also plot the results in Figure~\ref{fig:Rezs} together with
Planck constraints on $n_{\mathrm{s}}$ and $r$.

The large black dot in Fig.~\ref{fig:Rezs} corresponds to the ``pure''
$m^{2}\phi^{2}$ model with $\kappa=0$ and $N_{*}=60$, the same
as in Ref.~\cite{Planck_XXII}, while the smaller dot corresponds
to $N_{*}=50$. Blue curves show the effect of radiative corrections
with $h\gg g$, while red curves show this effect with $g\gg h$.
The solid line corresponds to the choice of the renormalization scale
$\mu=\phi_{*}$ while the dashed line corresponds to $\mu=\phi_{\e}$.
Inflaton coupling to fermions, i.e. negative $\kappa$, flattens the
potential, resulting in a smaller value of $r$ as compared to the
``pure'' $m^{2}\phi^{2}$ case. In contrast, a coupling to bosons
tends to steepen the potential, and thus increases the value of $r$.
For a very small coupling constant, radiative corrections have negligible
effect on the curvature of the potential. However, the number of efolds
$N_{*}$ decreases substantially. This can be seen in Figure~\ref{fig:Rezs}
as convergence of all the four curves on the joining line from $N_{*}=60$
point as they move upwards and towards the $N_{*}=50$ point.

As one can see in Figure~\ref{fig:Rezs}, different renormalization
scales result in different predictions for the CMB observables if
inflaton interactions are of the order of $g,\, h\sim10^{-3}$. The
gap between the blue solid and dashed curves in Figure~\ref{fig:Rezs}
is of the order $\Delta r\sim10^{-2}$, which is well within the sensitivity
of future missions such as CMBpol \cite{Baumann(2009)CMBPolInfl}
which has a planned precision of the same order, or PRISM \cite{Andre(2013)prism},
which is expected to reach the precision of $\Delta r\sim10^{-4}$.

We also plot the effect of radiative corrections on the running of
the spectral index $n'$ in Figure~\ref{fig:dn}. As one can see,
the difference in $n'$ between two renormalization schemes is of
the order of $10^{-4}$. Measurements with such a precision will certainly
be a major challenge for future missions, which cannot be achieved
by CMB observations alone; rather, one needs to probe the spectral
index on a much larger range of wavelengths. Such range can be measured
by combining CMB and Large Scale Structure observations, of which
the 21cm experiments may offer the best prospects towards this aim.
The Square Kilometer Array (SKA) \cite{SKAhomepage} will be able
to achieve the accuracy of a $\mathrm{few}\times10^{-3}$ for the
spectral running, while a more futuristic experiment called Fast Fourier
Transform Telescope (FFTT) \cite{Tegmark(2009)FFTT} could push this
limit down by one more order of magnitude \cite{Barger(2008)21cm,Adshead(2011)running}.

\begin{figure}
\centering{}\includegraphics[scale=0.5]{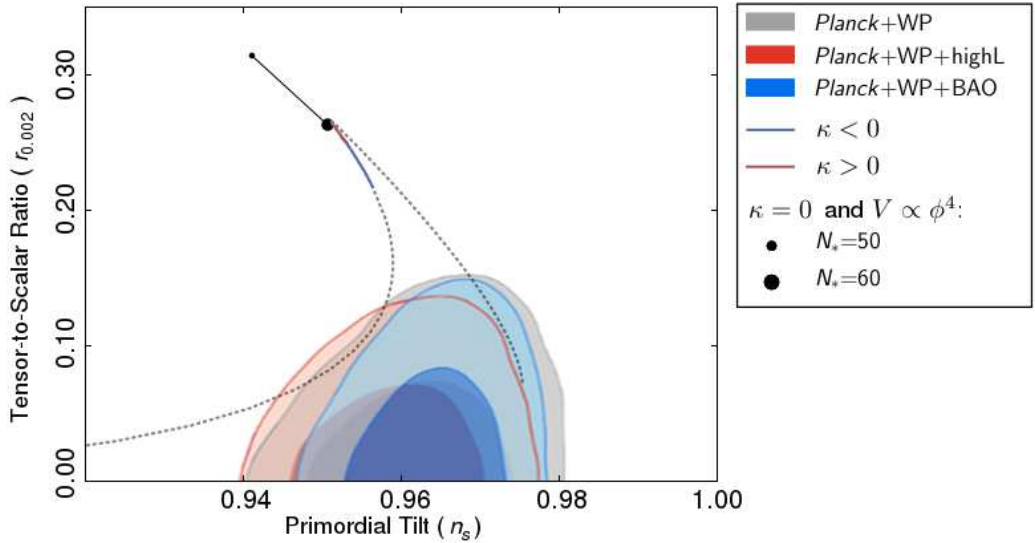}\caption{\label{fig:phi4}Planck constraints on the spectral index $n_{\mathrm{s}}$
and the tensor-to-scalar ratio $r$ \cite{Planck_XXII}. Black dots
correspond to the $\phi^{4}$ model with $\kappa=0$. The blue curve
correspond to the dominant decay chanel $\phi\rightarrow\psi\bar{\psi}$
($\kappa<0$) and the red one to $\phi\rightarrow\chi\chi$ ($\kappa>0$).
Solid curves show the effect of one loop radiative corrections, which
are subdominant to the tree level part of the potential, while gray
dotted curves show the effect of large $\left|\kappa\right|$ values,
such that radiative corrections dominate at least some of the inflationary
part of the potential.}
\end{figure}

\subsection{The Case of $n=4$}

The case of the $\lambda\phi^{4}$ potential is less interesting than
the $m^{2}\phi^{2}$ case. For the $\lambda\phi^{4}$ model the spectral
index $n_{s}$ and the tensor-to-scalar ratio $r$ do not depend on
the choice of the renormalization scale $\mu$ so that the situation
is much more straightforward. We plot $n_{s}$ and $r$ including
the one loop radiative corrections in Figure~\ref{fig:phi4}. If
we require that the radiative corrections remain smaller than the
tree level, the resulting modifications are still well out of the
Planck $2\sigma$ contours. If one were to extend the results into
the large $\left|\kappa\right|$ region (gray dotted curves in Figure~\ref{fig:phi4}),
where the radiative corrections would dominate at least some part
of the potential, one could meet the Planck constraints. However,
then one should worry about the role of the higher order corrections
so that obviously such a result cannot be trusted.

\section{Conclusions}

The precision with which observable inflationary parameters are measured
increased substantially over the last decade, culminating in the most
resent results from the Planck satellite. Indeed, the data from the
Planck satellite made it possible to exclude monomial $\phi^{4}$
and $\phi^{3}$ models of inflation with a high degree of confidence,
while $\phi^{2}$ models are on the verge of allowed region. This,
however, applies only to the tree-level potential, which neglects
the effects of inflaton interactions with other fields. Such interactions
are necessary in any realistic model of inflation for the inflaton
to reheat the universe into radiation dominated phase. These interactions,
however, modify the potential of the inflaton by introducing loop
corrections. In effect, the Planck constraints apply only to toy models,
and the obvious question then is, what are the constraints on (semi)realistic
models?

To the lowest order the modification of the inflaton potential is
a Coleman-Weinberg type correction as given in Eq.~\eqref{V-eff},
which is parametrized by the renormalization scale $\mu$. In particle
physics one fixes $\mu$ by determining physical masses and coupling
constants of fields from measurements of interaction amplitudes at
a given energy scale. For the inflaton, however, no observation exists
which would give independent determination of the potential parameters.

In this work we show how different choices of the renormalization
scale lead to different predictions for observable inflationary parameters.
We ignore corrections induced by the curvature and treat the inflaton
itself as a classical background field. We believe that such an approach,
although not completely without problems, will serve as a useful illustration
of the role of radiative corrections in modifying the naive observational
model constraints.

To give quantitative results we consider monomial chaotic type inflaton
potentials. Radiative corrections change the slope and curvature of
tree level potential, which in turn affects the predicted values of
the spectral index $n_{\mathrm{s}}$, its running $n'$ and tensor-to-scalar
ratio $r$. The effect of different choices of renormalization scale
$\mu$ is demonstrated by choosing $\mu=\phi_{*}$ and $\mu=\phi_{\e}$,
where $\phi_{*}$ and $\phi_{\e}$ are inflaton values when the pivot
scale exits the horizon and at the end of inflation respectively.
The results for the quadratic $\phi^{2}$ potential are summarized
in Figure~\ref{fig:Rezs} and some numerical parameter values are
given in Table~\ref{tab:rez}. Since the renormalization scale is
free, we end up with a family of models, each having a slightly different
slope of the potential, which could in principle be further constrained
by e.g. measuring the running of the spectral index. Meanwhile, we
can conclude that within the family of semirealistic chaotic $\phi^{2}$
models, some may be ruled out by Planck whereas some, and in particular
those in which the inflaton decays predominantly into fermions, remain
perfectly viable.

\section*{Acknowledgements}

KE is supported by the Academy of Finland grant 1218322; MK is supported
by the Academy of Finland grant 1263714.

\bibliographystyle{JHEP}
\bibliography{rCorrections.v03a}

\end{document}